Ferromagnetism at 300 K in spin-coated anatase and rutile $Ti_{0.95}Fe_{0.05}O_2$ films

R. Suryanarayanan[1*], V.M. Naik[2], P. Kharel[1], P. Talgala[1] and R. Naik[1]

[1]Department of Physics and Astronomy, Wayne State University, 666 W Hancock, Detroit, MI 48201, USA

[2]Department of Natural Sciences, University of Michigan-Dearborn, Dearborn, MI, 48128, USA

Thin films of $Ti_{1-x}Fe_xO_2$ (x=0 and 0.05) have been prepared on sapphire substrates by spin-on technique starting from metal organic precursors. When heat treated in air at 550 and 700 °C respectively, these films present pure anatase and rutile structures as shown both by X-ray diffraction and Raman spectroscopy. Optical absorption indicate a high degree of transparency in the visible region. Such films show a very small magnetic moment at 300 K. However, when the anatase and the rutile films are annealed in a vacuum of $1 \times 10^{-5}$ Torr at 500 °C and 600 °C respectively, the magnetic moment, at 300 K, is strongly enhanced reaching 0.46 $\mu_B$/Fe for the anatase sample and 0.48 $\mu_B$/Fe for the rutile one. The ferromagnetic Curie temperature of these samples is above 350 K.

Keywords: ferromagnetic films ; oxides, spin-coating, chemical, Raman effect, optical properties-

PACS: 75.60; 78.20.-e; 78.30.-j; 77.84. Bw

[*]Permanent address: LPCES, UMR8648, ICMMO, Université Paris-Sud, 91405 Orsay, France. E mail: sury39@yahoo.com

Transparent ferromagnetic semiconductors with ferromagnetic Curie temperature, $T_c>300$ K, are expected to play a vital role in the development of spintronics [1]. Earlier reports based on III-V semiconductors such as GaAs, doped with Mn, revealed that $T_c$ hardly exceeded 150 K rendering them unsuitable for viable room temperature applications [2]. Since then, the attention has turned towards the oxide semiconductors. Matsumoto et al., [3] was the first to report the occurrence of ferromagnetism with $T_c>300$ K in Co doped $TiO_2$ films prepared by pulsed laser deposition (PLD). Following this, several reports have appeared on the occurrence of ferromagnetism above room temperature in other materials such as $TiO_2$ doped with Fe [4,5], ZnO doped with Co [6,7], GaN with Mn [8], AlN doped with Cr [9] and $SnO_2$ doped with Co [10] or with Fe [11]. The ferromagnetic samples examined were prepared using different preparative techniques such as oxygen-plasma-assisted molecular-beam epitaxy (MBE) [12], sputtering [13], ion implantation [14], laser MBE [15] as well as PLD [5,9,10]. Though all these samples did show unambiguous signs of ferromagnetic behavior at 300 K or above, it was difficult to exclude the possibility of formation of clusters of dopants and thus attribute the observed magnetism entirely to 3d element substituted in different hosts. Several arguments, for and against the formation of clusters, were proposed by these authors. It should be further noted that the double exchange mechanism [16], which is known to account for the ferromagnetism in mixed valence manganites, may not be applicable in the above materials. Another mechanism in vogue involves ferromagnetic coupling via spin-polarized $p$-band holes to account for the ferromagnetic behavior in $Ga_{1-x}Mn_xAs$ [17]. This model is also not expected to account for the ferromagnetic behavior observed in the 3d doped oxides or nitrides. However, carrier mediated ferromagnetism with $T_c>300$ K has been predicted to occur in Mn doped ZnO [18].

We have used a non-vacuum approach, known as spin-on technique, to prepare Fe substituted $TiO_2$ films. This is a well-known technique that has been used to prepare $TiO_2$ films starting from metalorganic precursors [19]. Some of the salient features of this technique are: low cost, simplicity, possibility of using starting materials of high purity and the possibility to obtain a homogeneous mixture of two cations in the liquid state. We report here on the successful preparation of 5% Fe substituted $TiO_2$ films crystallizing either in anatase or rutile structure and present X-ray diffraction (XRD), optical absorption, Raman spectra and magnetization measurements. Though the as prepared films showed a very weak ferromagnetic behavior at 300 K, the vacuum annealed films showed a strong enhancement in the magnetic moment of Fe.

The metalorganic precursors used were Ti-ethylhexoxide and Fe-ethylhexoxide. The solutions were thoroughly mixed in approximately 95 to 5 ratio (by volume) using an ultrasonic bath. A small amount of xylene was added to obtain optimal viscosity needed for spin-coating. A few drops of this mixed solution were placed on a sapphire substrate (c-axis oriented of 25 mm$^2$ area) which was spun at 5000 rpm for 15 sec. The sample was annealed in air at 550 $^o$C for one minute and the process was repeated 5 to 10 times to build up the layer thickness. The thickness of the sample used in the present study was 500 nm. Several such samples were prepared and were subjected to different heat treatments in air. The EDX showed that the concentration of Fe was about 5%. Films were further vacuum annealed under $1 \times 10^{-5}$ Torr at 500 $^o$C or 600 $^o$C.

The as prepared films with and without Fe heat treated at 550 $^o$C in air show a pure anatase phase whereas the films annealed at 700 $^o$C in air show a rutile structure, as revealed by XRD. For example, Fig. 1a and 1b show XRD pattern of undoped anatase and rutile TiO$_2$ films of ~1μm thickness. The films are polycrystalline with no preferred orientation. The different phases of the films were further confirmed by Raman spectra as shown in Figs. 2(a) and 3(a), which clearly exhibit all the characteristic phonon modes, expected for the two forms of TiO$_2$. These are in full agreement with the spectra observed in single crystalline forms of anatase and rutile -TiO$_2$ both of which belong to D$_{4h}$ space group [20, 21]. The broad band ~235 cm$^{-1}$ in rutile-TiO$_2$ (Fig. 3a) has been assigned as a combination line rather than a fundamental one phonon process [21]. The mode at 144 cm$^{-1}$ (B$_{1g}$) is expected to be very weak in pure rutile phase. The Raman spectra of Ti$_{0.95}$Fe$_{0.05}$O$_2$ (figs.2b and 3b) are identical to those of undoped TiO$_2$ films in respective forms indicating that the structures are retained after doping with Fe. This suggests that the Fe-dopants occupy the substitutional sites in the host lattice. Further, no noticeable changes in the spectra were seen due to the vacuum annealing (spectra not shown). The optical absorption spectra of anatase and rutile-Ti$_{0.95}$Fe$_{0.05}$O$_2$ films with and without vacuum annealing are shown in Figs. 4 and 5. There are no drastic changes in the optical absorption spectra after vacuum annealing. The observed intensity oscillations, especially in anatase film, in the wavelength region (λ) of 450 – 800 nm are due to thin film optical interference effects whereas the strong increase in the absorption for λ < 400 nm is due to electronic band gapabsorption.  Anatase- Ti$_{0.95}$Fe$_{0.05}$O$_2$ has an absorption edge ~350 nm, whereas the rutile-

$Ti_{0.95}Fe_{0.05}O_2$ has an absorption edge at around 375 nm. These optical absorption edges are very close to the ones observed for undoped $TiO_2$ samples. The observed absorption edges in the present work are consistent with the reported values in the literature for the two forms of $TiO_2$.

The magnetization (M) of the Fe substituted samples before and after vacuum annealing was measured as a function of temperature and magnetic field (H) using a Quantum Design superconducting quantum interference device (SQUID) magnetometer. M as a function of H (-5000 to 5000 Oe) at 300 K of the blank substrate was also measured in order to eliminate the substrate contribution. The data shown here thus represent the magnetization of the films. Figure 6 shows M as a function of H at 300 K of anatase sample before and after annealing. Several interesting features can be noted. First, at 300 K, the as prepared anatase form of Fe substituted $TiO_2$ film shows a very small magnetic moment (0.03 $\mu_B$/Fe at 2000 Oe). Second, the vacuum annealed sample at 500 $^o$C, shows a strong enhancement in the ferromagnetic properties, the magnetic moment reaching a value of 0.46 $\mu_B$/Fe at 2000 Oe. Also, the coercivity increases remarkably from almost zero to 600 Oe as a result of this annealing. The remanent magentization ($M_r$) at 300 K is 0.19 $\mu_B$/Fe. The M of the rutile sample is also very sensitive to the vacuum annealing treatment but the behavior is slightly different from that of the anatase sample(fig.7). The value of M of the as prepared rutile sample is ~ 0.09 $\mu_B$/Fe at 2000 Oe and increases to only 0.14 $\mu_B$/Fe after annealing at 500 $^o$C. However, annealing the sample at 600 $^o$C increases the magnetic moment to 0.48 $\mu_B$/Fe. A similar behavior was observed in the Co doped $TiO_2$ samples obtained by a similar spin-on technique [22]. The coercivity of the rutile sample also increases remarkably from a few Oe to 230 Oe. The value of $M_r$ is 0.12 $\mu_B$/Fe. Further, M as a function of temperature, measured in a field of 2000 Oe, in the case of both the anatse and the rutile samples, after annealing at 500 and 600 C respectively is fairly constant from 5 K to 350K (Figs. 6 and 7 inset). Since the temperature range of our magnetization measurements is limited to 350 K, we expect our samples to remain ferromagnetic at least up to 400 K.

In what follows, we present a short discussion of our data in comparison with those published earlier and further comment briefly on a recent model that has been proposed by Coey et al., [23] and Venkatesan et al., [24] to account for the occurrence of ferromagnetism in transparent semiconductors. To start with, we note that there has been very few reports on Fe doped $TiO_2$. These describe either rutile phase [4, 25] or a mixture of anatase and rutile phase [26]. On the contrary, we could obtain either the anatase phase or the rutile phase on the sapphire substrates just by selecting a proper heat treatment, as revealed unambiguously by Raman data. And further, the pure phases are retained after the heat treatment in vacuum. Next we comment on the magnetic moment of Fe obtained at 300 K. Wang et al.[4] have reported a value of about 2.3 $\mu_B$/Fe in the case of 6% of Fe in $TiO_2$ whereas Hong et al.[25] have obtained a value of 0.13 $\mu_B$/Fe in the case of 8 % Fe doped samples, though both these authors have used the same method of pulsed laser ablation to prepare the doped films. Our values lie in between these values. The different values reported could be due to various reasons. For example, a part of Fe can be present in the samples as clusters and some part can be located at the Ti sites. Further, the presence of oxygen defects may also influence the magnetic moment. Indeed, in our case, the heat treatment has a remarkable effect on the value of M. The fact that the magnetic moment and the coercivity of the samples exhibiting anatase phase is different from that observed for the rutile phase samples may also indicate that the nature of oxygen defects are different in these two different phases. It is interesting to note that in the case of Co doped ZnO films, the magnetic moment of Co strongly depended on the preparation conditions such as the partial pressure of oxygen showing the role played by oxygen deficiencies [24]. Next, we comment on the formation of magnetic clusters in these films. Murakami *et al*. [15] concluded from the EXAFS, XANES and XPS studies that their rutile films contained both Co metallic clusters and Co substituted into Ti site. Such studies have not been carried out on Fe doped samples. We may note, however, magnetic force microscope studies carried out in the case of Fe doped $SnO_2$ samples [23] and V doped $TiO_2$ samples [27], do not seem to support the formation of nanometer-sized clusters on the surface and thus the observed magnetic properties were taken to represent intrinsic behavior of the samples. Since there is no exchange mechanism which could account for the observed data at low doping levels, Coey et al., [23] have proposed a model where the exchange is mediated by carriers in a spin-split impurity band derived from extended donor orbitals. They point out that the oxygen vacancies play an important role. Though the details have not been worked out yet, this model is very appealing to us since annealing of our films in vacuum, could possibly create oxygen vacancies that could influence the magnetic properties in this diluted system. In the present case, it may also induce another valence state of Ti, assuming Fe is in a valence state of 3. Detailed studies on the effect of annealing on the magnetic properties and the microstructure of the $TiO_2$ doped with Fe and other elements seem to be highly desirable in order to understand further this interesting aspect. Such studies are expected to throw further light on the role played by oxygen defects on the magnetic properties. We end this discussion by commenting on a recent theoretical work by Bergqvist *et al*. [28]. By noting that the magnetic properties on oxide magnetic semiconductors depend very much on the preparation conditions, these authors propose a model based on magnetic percolation. They point out that the random distribution of magnetic atoms in the host lattice plays a crucial role and underline the importance of the localization of exchange interactions that have a strong directional dependence.

In summary, by using a low cost spin-on technique, we have successfully prepared Fe(≈5%) substituted $TiO_2$ films on sapphire substrates both in the anatase and in the rutile forms that show room temperature ferromagnetism. The magnetic properties are found to be enhanced as a result of annealing in vacuum. Upon annealing at 500 $^o$C, the magnetic moment and the coercivity, at 300 K, were found respectively to be 0.46 $\mu_B$/Fe and 600 Oe for the anatase sample and 0.48 $\mu_B$/Fe and 230 Oe for the rutile one. Our data, though cannot rule out the presence of Fe clusters, point out the important role played by oxygen defects.

Acknowledgment

One of the authors (VMN) acknowledges the support from the Office of Vice President for Research at the University of Michigan-Ann Arbor.

Figure Captions

Fig. 1  XRD θ–2θ scans of (a) anatase and (b) rutile TiO$_2$ films

Fig. 2  Raman spectra of anatase (a) TiO$_2$ and (b) Ti$_{0.95}$Fe$_{0.05}$O$_2$ films. The asterisks represent the peaks originating from the sapphire substrates.

Fig. 3  Raman spectra of rutile (a) TiO$_2$ and (b) Ti$_{0.95}$Fe$_{0.05}$O$_2$ films. The band marked with an asterisk is a combination band.

Fig. 4  Optical absorption spectra of anatase Ti$_{0.95}$Fe$_{0.05}$O$_2$ film before and after vacuum annealing.

Fig. 5  Optical absorption spectra of rutile Ti$_{0.95}$Fe$_{0.05}$O$_2$ film before and after vacuum annealing.

Fig. 6  Magnetic Moment vs Magnetic field at 300 K for anatase Ti$_{0.95}$Fe$_{0.05}$O$_2$ film (symbols • and o represent the data before and after vacuum annealing at 500 $^o$C). The inset shows the Magnetic Moment vs Temperature data in a field of 2000 Oe.

Fig. 7  Magnetic Moment vs Magnetic field at 300 K for rutile Ti$_{0.95}$Fe$_{0.05}$O$_2$ film (symbols • and o represent the data before and after vacuum annealing at 600 $^o$C) The inset shows the corresponding magnetic moment vs temperature data in a field of 2000 Oe.

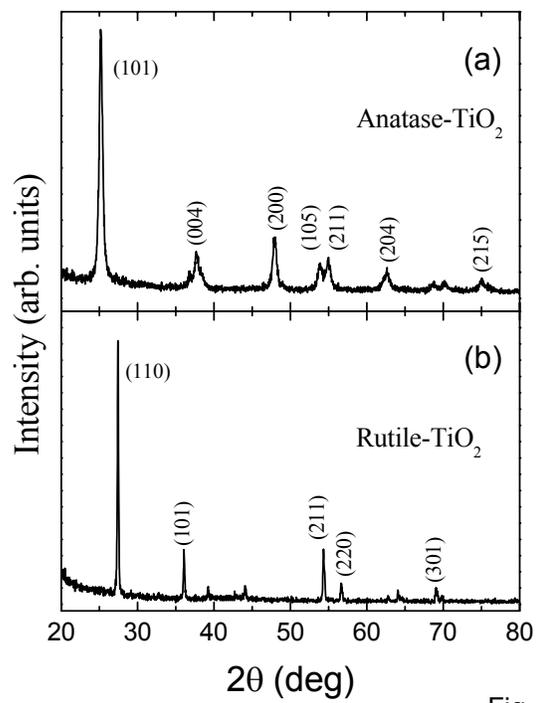

Fig. 1

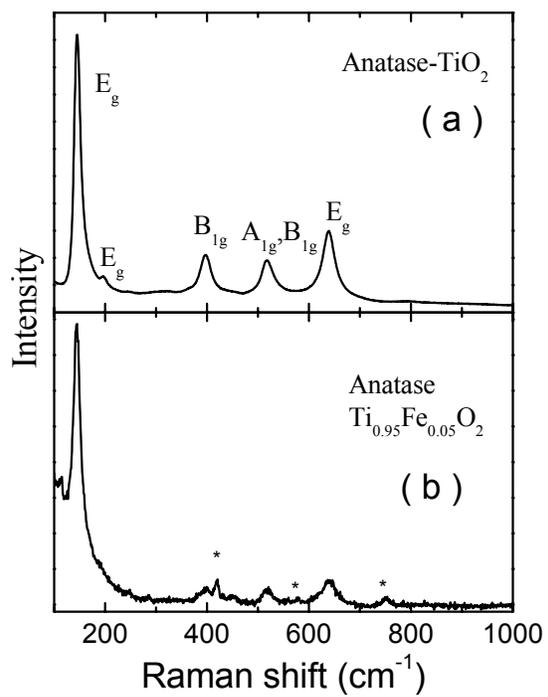

Fig.2

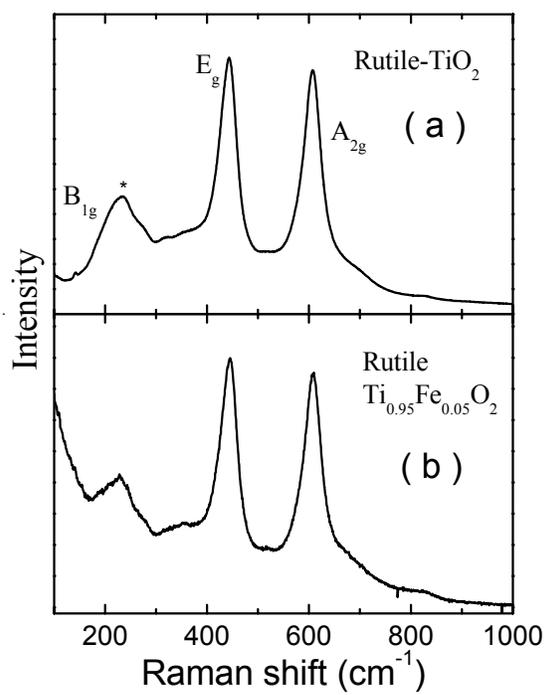

Fig.3

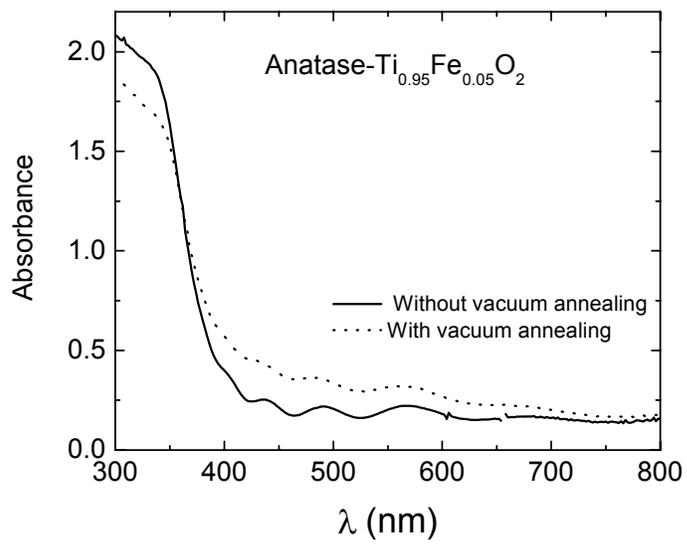

Fig.4

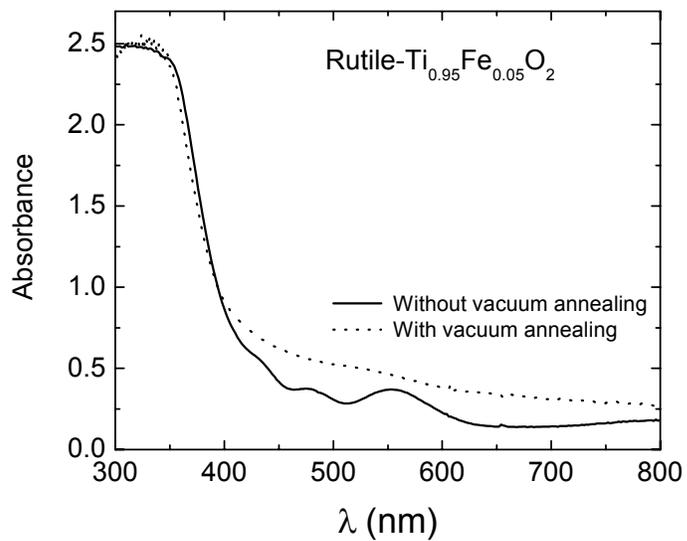

Fig.5

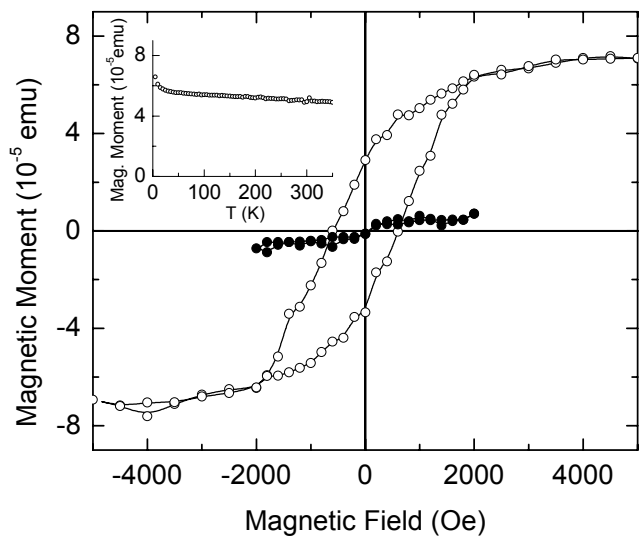

Fig.6

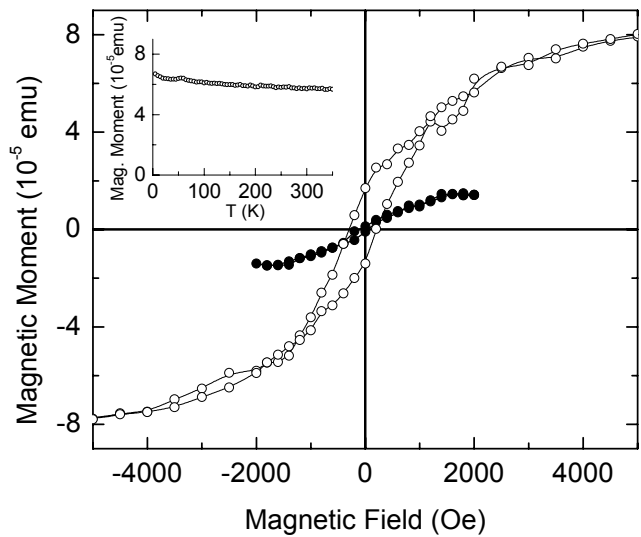

Fig.7